\documentclass[fleqn,usenatbib]{mnras}

\usepackage{newtxtext,newtxmath}

\usepackage[T1]{fontenc}

\DeclareRobustCommand{\VAN}[3]{#2}
\let\VANthebibliography\thebibliography
\def\thebibliography{\DeclareRobustCommand{\VAN}[3]{##3}\VANthebibliography}

\usepackage{graphicx}	
\usepackage{amsmath}	
\usepackage{longtable}  
\usepackage{multirow}
\usepackage{manyfoot}
\usepackage{lipsum}
\usepackage{mathtools}
\usepackage{mathrsfs}
\usepackage{placeins}
\usepackage{booktabs}
\usepackage{xcolor}
\usepackage{lscape}
\usepackage{rotating}
\usepackage{multirow}
\usepackage{graphicx}

\definecolor{mygreen}{rgb}{0,0.502,0}
\definecolor{myblue}{rgb}{0.07058824, 0.07058824, 0.5372549}
\hypersetup{linkcolor=mygreen,citecolor=mygreen,urlcolor=mygreen}

\graphicspath{{figs/}}

\usepackage{xspace}

\newcommand*{\XPSI}{\texttt{X-PSI}\xspace}
\newcommand*{\NICER}{NICER\xspace}

\newcommand*{\PyMultiNest}{\textsc{PyMultiNest}\xspace}
\newcommand*{\MultiNest}{\textsc{MultiNest}\xspace}

\title[Polarized pulses]{Modelling polarized X-ray pulses from accreting millisecond pulsars with X-PSI, using different hot spot locations and shapes}

\author[Salmi~et~al.]{Tuomo~Salmi,$^{1,2}$\thanks{E-mail: tuomo.salmi@helsinki.fi}
Bas~Dorsman,$^{1}$
Anna~L.~Watts,$^{1}$
Anna~Bobrikova,$^{3}$
Alessandro~Di~Marco,$^{4}$\newauthor
Vladislav~Loktev,$^{2,3}$
Alessandro~Papitto,$^{5}$
Maura~Pilia,$^{6}$
Juri~Poutanen,$^{3,7}$ and
John~Rankin$^{4,8}$
\\
$^{1}$Anton Pannekoek Institute for Astronomy, University of Amsterdam, Science Park 904, 1098XH Amsterdam, the Netherlands\\
$^{2}$Department of Physics, P.O. Box 64, FI-00014 University of Helsinki, Finland\\
$^{3}$Department of Physics and Astronomy, FI-20014 University of Turku, Finland\\
$^{4}$INAF Istituto di Astrofisica e Planetologia Spaziali, Via del Fosso del Cavaliere 100, 00133 Roma, Italy\\
$^{5}$INAF Osservatorio Astronomico di Roma, Via Frascati 33, 00078 Monte Porzio Catone (RM), Italy\\
$^{6}$INAF Osservatorio Astronomico di Cagliari, Via della Scienza 5, 09047 Selargius (CA), Italy\\
$^{7}$Space Research Institute, Russian Academy of Sciences, Profsoyuznaya 84/32, Moscow 117997, Russia\\
$^{8}$INAF Osservatorio Astronomico di Brera, Via E. Bianchi 46, 23807 Merate, Italy
\\
}

\date{Accepted 2025 March 10. Received 2025 March 6; in original form 2025 January 21}

\pubyear{\the\year{}}

\begin{document}
\label{firstpage}
\pagerange{\pageref{firstpage}--\pageref{lastpage}}
\maketitle

\begin{abstract}
We present an analysis of polarized X-ray pulses based on simulated data for accreting millisecond pulsars (AMPs). 
We used the open-source X-ray Pulse Simulation and Inference code (previously applied to NICER observations), which we upgraded to allow polarization analysis.
We provide estimates of how well neutron star (NS) parameters can be constrained for the Imaging X-ray Polarimetry Explorer (IXPE) and find that strong limits on the hot region geometries can be hard to obtain if the emitting hot region is large and the number of polarized photons relatively small.
However, if the star is bright enough and the hot regions are small and located so that polarization degree is higher, the observer inclination and hot spot colatitude can be constrained to a precision of within a few degrees.
We also found that the shape of the hot region, whether a circle or a ring, cannot be distinguished in our most optimistic scenario. 
Nevertheless, future X-ray polarization missions are expected to improve the constraints, and already the recent AMP polarization detections by IXPE should help to infer the NS mass and radius when combined with modelling of X-ray pulse data sets that do not contain polarization information.

\end{abstract}

\begin{keywords}
dense matter -- polarization -- methods: numerical -- stars: neutron -- techniques: polarimetric -- X-rays: binaries 
\end{keywords}
 
\section{Introduction}
\label{sec:intro}
 
X-ray observations of neutron stars (NSs) can be used to probe both the interior and exterior properties of the NSs.
The equation of state, i.e., the pressure--density relation of the high-density NS core can be determined by measuring masses and radii from a set of NSs \citep[see e.g.][]{Lattimer12ARNPS,Baym2018}. 
One method to infer masses and radii is to model the energy-resolved X-ray pulses originating from hot regions on top of rapidly rotating NSs (millisecond pulsars), and exploit the relativistic effects on the spectra and the shape of the pulses \citep[see e.g.][]{PFC83,ML98,PG03,MLC07,SNP18,BLM_nicer19}.
This method allows also the exploration of the emission physics and magnetic field properties of these systems \citep[see e.g.][]{Bilous19,Kalapotharakos2021,Salmi2023,vinciguerra2024bravo}.

Even though pulse profile modelling can provide useful constraints on NS parameters, it often faces a challenge in breaking the degeneracy between the pulsar geometry (especially inclination angle and magnetic obliquity) and the mass and radius. 
However, this degeneracy can be mitigated by observing and modelling the variations in X-ray polarization with pulsar phase \citep{VP04,Poutanen10}. 
The polarization itself is only weakly dependent on mass and radius \citep{poutanen20,LSNP20}, but it can help to constrain the pulsar geometry.
Observable polarization is expected particularly in the case of accretion-powered millisecond pulsars \citep[AMPs;][]{patruno21,disalvo22}, where the emission is polarized due to Compton scattering of thermal seed photons by hot electrons in the accretion column \citep{PS96,SLK21,Bobrikova2023}.
Recently, polarized X-rays were discovered by the Imaging X-ray Polarimeter Explorer \citep[IXPE;][]{Weisskopf2022} during the outburst of the AMP SRGA J144459.2$-$604207 \citep{papitto25}.
Further modelling of these data and observations with future satellites, such as the enhanced X-ray Timing and Polarimetry mission \citep[][]{EXTP,dmatter_extp}, are expected to provide increasingly better geometry constraints.

Previously, polarized emission models for AMPs were developed using either Thomson scattering \citep{ST85,VP04,SLK21} or more accurate Compton scattering approaches \citep{PS96,Bobrikova2023}.
NS parameter constraints based on synthetic IXPE observations were also investigated in \citet{SLK21} and \citet{Bobrikova2023}.
In this work, these simulations are extended to allow more free parameters (including mass and radius), additional choices for the input parameter values \citep[matching new SAX~J1808.4$-$3658-like synthetic NICER data analysis of][]{dorsman25}, and more complex spot shapes than uniformly emitting circles. 
The modelling is also performed this time using the upgraded open-source X-ray Pulse Simulation and Inference code (\XPSI, \citealt{xpsi}), which has been used so far to model X-ray pulses observed by NICER \citep[see e.g.][]{RWB_nicer19,Riley2021,choudhury24,Salmi2024,Salmi2024b,vinciguerra2024bravo} and RXTE \citep{kini24b} without polarization data. 
This implementation allows straightforward joint analysis of pulse profile data sets with and without polarization information (taken by different instruments).

The remainder of this paper is structured as follows.
In Section~\ref{sec:model}, we present the methods.  
In Section~\ref{sec:results}, we present the results. 
We discuss the implications of our results for future analysis in Section~\ref{sec:discussion} and conclude in Section~\ref{sec:conclusions}.

\section{Modelling procedure}\label{sec:model}

\begin{table*}
    \caption{Model parameters and values chosen for each scenario. The priors used for sampling are also given, where $\mathcal{U}$ means uniform distribution with lower and upper bound. Unchanged parameters are indicated with a dash.}\label{table:input_params}
    \begin{tabular}{lllcccc} \hline
    Parameter (Unit) & Description & Prior density & \multicolumn{2}{c}{Input value}  \\
     & & & Case A & Case B & Case C  & Case D \\ \hline
    $D$ (kpc) & Distance & fixed & 2.7 & - & - & -\\
    $M$ (M$_\odot$) & Mass & $\mathcal{U}(1,3)$ & 1.40 & - & - & -\\
    $R_\mathrm{eq}$ (km) & Equatorial radius & $\mathcal{U}(3r_{\mathrm{G}}(1),16)^{\mathrm{a}}$ & 11.0 &- & - & -\\
    $i$ (deg) & Inclination & $\cos i \sim \mathcal{U}(0, 1)$ & 40 & 80 & 10 & 10 \\    
    $f$ (Hz) & Pulsar spin frequency & fixed & 401 & - & - & -\\
    $\phi_{0}$ (cycles) & Phase zero& $\mathcal{U}(-0.5,0.5)$ & 0 & - & - & -\\
    $\theta_{\mathrm{p}}$ (deg) & Colatitude & $\cos \theta \sim \mathcal{U}(-1, 1)$ & 10.3 & 10.3 & 105 & 105\\    
    $\zeta_{\mathrm{p}}$ (deg) & Primary angular radius& $\mathcal{U}(0, 90)$ & 89.9 & 30.0 & 1.0 & 10.0\\
    $\zeta_{\mathrm{o}}$ (deg) & Mask angular radius & $\mathcal{U}(0, 90)^{\mathrm{b}}$ or fixed$^{\mathrm{c}}$ & 0.0 & 0.0 & 0.0 & 9.0\\    
    $\chi_{0}$ (deg) & Spin axis position angle & $\mathcal{U}(-90, 90)$ & 0 & - & - & -\\    
    $T_\mathrm{seed}$ (keV) & Seed photon temperature & $\mathcal{U}(0.5,1.5)$ & 0.52 & 1.28 & 1.28 & 1.28\\
    $T_\mathrm{e}$ (keV) & Electron slab temperature & $\mathcal{U}(20,102)$ & 37.0 & 51.1 & 51.1 & 51.1\\
    $\tau$ (-) & Thomson optical depth & $\mathcal{U}(0.5,3.5)$ & 1.5 & 2.0 & 2.0 & 2.0\\
    $N_\mathrm{H}$ ($10^{21}{\mathrm{cm}^{-2}}$) & ISM column density & $\mathcal{U}(0,10)$ & 1.17 & - & - & -\\\hline
    \end{tabular}\\ 
    \vspace{2 mm}
    \begin{flushleft}
    \footnotesize{$^{\mathrm{a}}$ $r_{\mathrm{G}}(1)$ is the gravitational radius of $M=1$ M$_\odot$. The prior is also modified by the compactness condition, requiring that $R_{\textrm{polar}}/r_{\rm g}(M)>3$.}\\
    \footnotesize{$^{\mathrm{b}}$ The prior is also modified by requiring that $\zeta_{\mathrm{o}} < \zeta_{\mathrm{p}}$.}\\
    \footnotesize{$^{\mathrm{c}}$ Fixed in Cases A, B, and C.}    
    \end{flushleft}
\end{table*}

\subsection{Polarized X-ray pulse profile modelling}

We model the emission from the NS surface to the observer using the pulse profile modelling technique as e.g. in \citet{BLM_nicer19}. 
We additionally calculate the polarization angle (PA) transportation accounting for  relativistic effects \citep{VP04,poutanen20} including the oblate shape of the star \citep{LSNP20}.
As in \citet{dorsman25}, we adopt the emission model from \citet{Bobrikova2023}, which describes how the photon intensity on the NS surface depends on the cosine of the emission angle $\mu$ (between the ray and the local normal) and on the photon energy $E$. 
In this model photons are Comptonized when they travel through a plane-parallel slab of hot electrons (heated by the accretion process), and the outcoming intensity depends on three parameters: seed photon temperature $T_{\rm seed}$, electron temperature $T_{\rm e}$, and Thomson optical thickness $\tau$ of the slab.
To speed up the calculation, we use pre-computed tables and an interpolation technique similar to that of \citet{dorsman25}.
The only difference is that in addition to Stokes $I(\mu,E)$, we interpolate the pre-computed values for Stokes $Q(\mu,E)$, which together define the PA and polarization degree (PD) in the frame of the NS surface.
The Stokes $U$ parameter is zero in this frame (at any specific point on the surface) due to the assumed azimuthal symmetry of the radiation field around the surface normal. This means that the polarization vector has to lie in the plane containing the normal or be perpendicular to it, and thus the PA must be either $0$ or $\pi/2$. 
The azimuthal symmetry is expected for AMPs, where the magnetic field does not affect the polarization properties.
In addition, the Stokes $V$ parameter is zero because Compton scattering is not a source of circular polarization.

After rotating the PA to the frame of the observer, we obtain the observed Stokes fluxes $F_{I}$, $F_{Q}$, and $F_{U}$ (see e.g. equation 21 in \citealt{poutanen20}).
As in the previous \XPSI analyses, we integrate the observed fluxes -- this time all the Stokes components -- over the hot region surface area, which is discretized with a regular mesh of points in colatitude and azimuth about the stellar spin axis (see section B.5 in \citealt{BLM_nicer19} for details).
The shape of the spot is either assumed to be a circle, or is constructed using multiple overlapping circles.
In this paper, we consider two different hot region cases: a single-temperature (\texttt{ST}) and a concentric single-temperature (\texttt{CST}) model.
The latter model corresponds to a ring with an additional parameter $\zeta_{\mathrm{o}}$ defining the angular radius of the “omitting” circle that masks the emitting circle.

Unlike in \citet{dorsman25}, we do not account for the photons emitted from an accretion disc.
This would mostly affect the X-rays that are below 2 keV (the lowest IXPE energies).
Using a similar disc model and parameters to \citet{dorsman25}, we verified that the disc contribution to the flux at 2 keV is only a few per cent, which is much smaller than the polarization measurement uncertainty even in our most optimistic case. 
However, especially for future polarimetric instruments, this could be studied in more detail by modelling the polarization coming from the accretion disc using the technique presented in \citet{loktev2022}.

The attenuation of the X-ray flux due to the interstellar medium also affects mostly the energies below the IXPE energy range. 
However, since accounting for it is straightforward and relatively inexpensive, we use the same \texttt{tbnew} model \citep{wilms2000} as in \citet{dorsman25}, with neutral hydrogen column density $N_{\mathrm{H}}$ as a free parameter.

Besides the parameters used in non-polarized pulse profile modelling, we also account for the spin axis position angle $\chi_{0}$, which is the angle between the observer north and the projection of the rotation axis on the plane of the sky.
The final modelled Stokes parameters are therefore
\begin{eqnarray} \label{eq:chi_def_}
F^{\mathrm{mod}}_{I} &=& F_{I}, \nonumber \\
F^{\mathrm{mod}}_{Q} &=& F_{Q}\cos(2 \chi_{0})-F_{U}\sin(2\chi_{0}),  \\
F^{\mathrm{mod}}_{U} &=& F_{Q}\sin(2 \chi_{0})+F_{U}\cos(2\chi_{0}).  \nonumber
\end{eqnarray}

For the pulse simulations and inference runs, we used the \XPSI \citep{xpsi}\footnote{\url{https://github.com/xpsi-group/xpsi}}  code \texttt{v2.2.3}, where the support of polarimetry and the new atmosphere interpolation method were implemented.
\XPSI \texttt{v3.0.0} was used for producing the figures.

\subsection{Input model scenarios}\label{sec:input_models}

To generate a synthetic data set, we consider four different input model scenarios (see Table \ref{table:input_params} for the exact parameter values in each):
\begin{itemize}
\item Scenario A: one \texttt{ST} hot spot with parameters similar to those in Scenario~A in \citet{dorsman25}, i.e. a large spot with the viewing geometry and NS emission model parameters selected based on a fit of 2019 \NICER data of SAX~J1808.4$-$3658. 
\item Scenario B: similar to Scenario~B in \citet{dorsman25}, i.e., a \texttt{ST} hot spot with parameters selected based mostly on Scenario~A, but with a smaller and hotter spot to produce at least $\sim 2$ per cent PD and having roughly the same number of X-ray counts in the IXPE energy band as in the previous scenario (within a factor of 2). 
Electron temperature $T_{\mathrm{e}}$ and optical depth $\tau$ were also slightly increased to boost the PD.
Most photons have higher energy than in Scenario A due to the higher seed photon temperature $T_{\mathrm{seed}}$.
\item Scenario C: one \texttt{ST} hot spot with parameters selected to further improve the polarization detection significance by scaling the source flux to 100 mCrab and using a similar hot spot location and viewing angle as in set 4 from \citet{Bobrikova2023}.
\item Scenario D: similar to Scenario C, but assuming a \texttt{CST} (ring-like) hot spot shape with 9\degr\ inner and 10\degr\  outer angular radius.
\end{itemize}

\subsection{Synthetic data production}

We used the \texttt{ixpeobssim} software \citep[][versions 30 and 31]{ixpeobssim}\footnote{\url{https://github.com/lucabaldini/ixpeobssim}} coupled with \XPSI to produce the synthetic polarization data for IXPE.
The effective observing time of the source was assumed to be 600 ks, and the flux of the source was calculated directly from the model parameters in Scenarios A and B (being around 10--20 mCrab), and scaled to 100 mCrab in Scenarios C and D.
The latter corresponds to roughly SAX~J1808.4$-$3658 during its typical peak luminosity.
We used the \texttt{pcube} algorithm to bin the data into 10 phase bins and into one single energy bin (between $2$ and $8$ keV).
The resulting normalized Stokes $q=Q/I$ and $u=U/I$ data sets are presented in Section \ref{sec:simulated_data} and modelled in Section \ref{sec:parameter_constraints}.
For some exploratory analyses, we also binned the Stokes $I$ data into 20 phase bins and included the corresponding $i_{\mathrm{N}}=I/I_{\mathrm{max}}$ in the modelling.

\subsection{Posterior computation}

As in \citet{SLK21} and \citet{Bobrikova2023}, we calculate the likelihood of the simulated normalized Stokes $q$ and $u$ data by comparing them to the corresponding modelled Stokes parameters $q^{\mathrm{mod}} = F_{Q}^{\mathrm{mod}}/F_{I}^{\mathrm{mod}}$ and $u^{\mathrm{mod}} = F_{U}^{\mathrm{mod}}/F_{I}^{\mathrm{mod}}$.
As before, we assume that $q^{\mathrm{mod}}$ and $u^{\mathrm{mod}}$ are uncorrelated and normally distributed around $q$ and $u$ with measured errors as the standard deviation. 
This time, in Scenario B we additionally explore the effect of including the likelihood of the simulated Stokes $i_{\mathrm{N}}$ data by comparing $i_{\mathrm{N}}$ to $i^{\mathrm{mod}} = F_{I}^{\mathrm{mod}}/F_{I,\mathrm{max}}^{\mathrm{mod}}$ and assuming a Gaussian likelihood function.

When fitting the data, we kept most of the model parameters free using the prior distributions presented in Table \ref{table:input_params}.
The posterior distributions of the parameters were obtained using \XPSI coupled with  \MultiNest \citep{MultiNest_2008,multinest09,FHCP2019} through \PyMultiNest \citep{PyMultiNest}.
For the sampler settings we used $0.1$ sampling efficiency and $10^{4}$ live points.
In some cases, we tested that using 4000 live points generated similar results, which indicates the settings used are more than sufficient.

\section{Results}\label{sec:results}

{
\begin{figure*}
\centering
\includegraphics[
width=0.49\textwidth]
{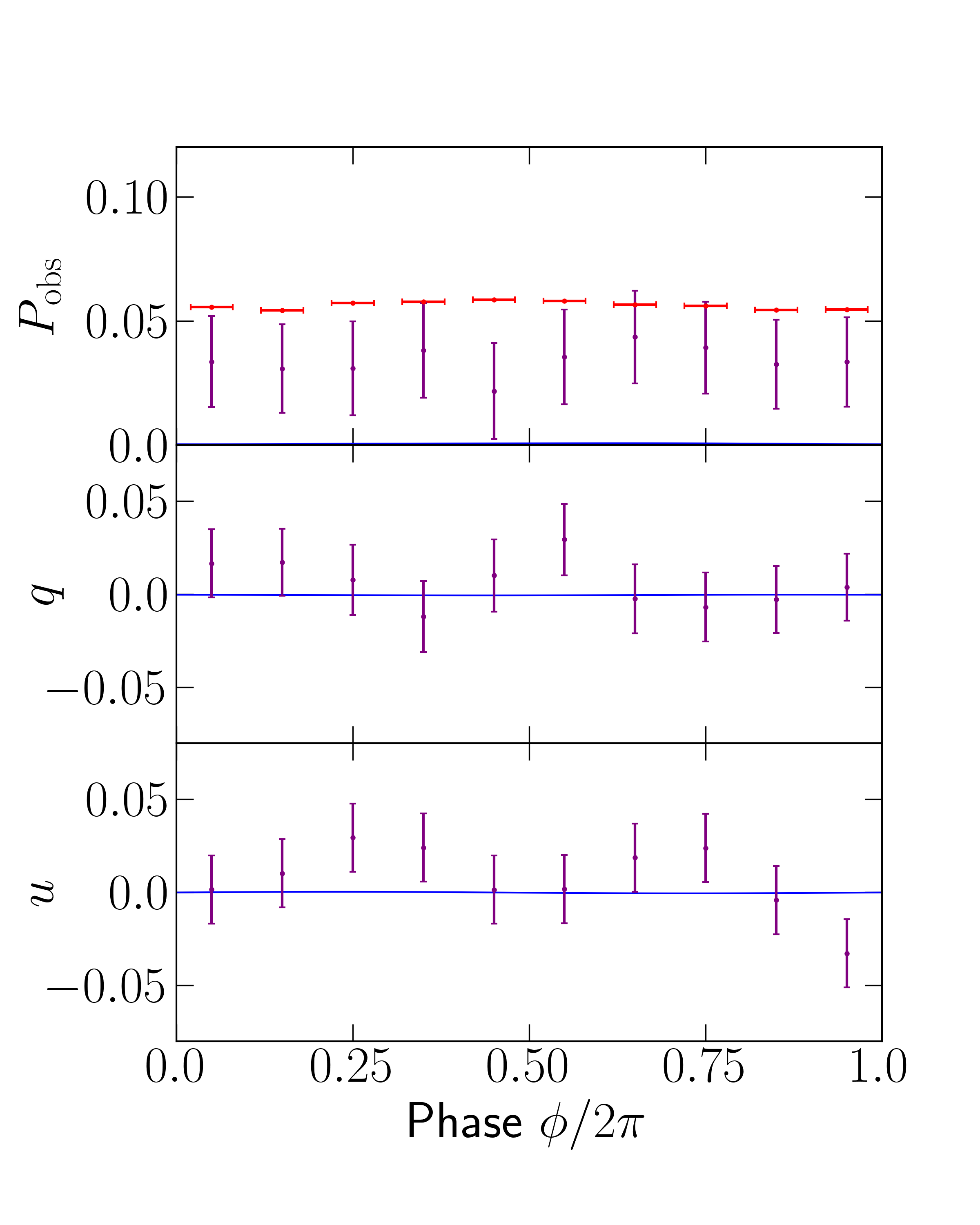}
\includegraphics[
width=0.49\textwidth]
{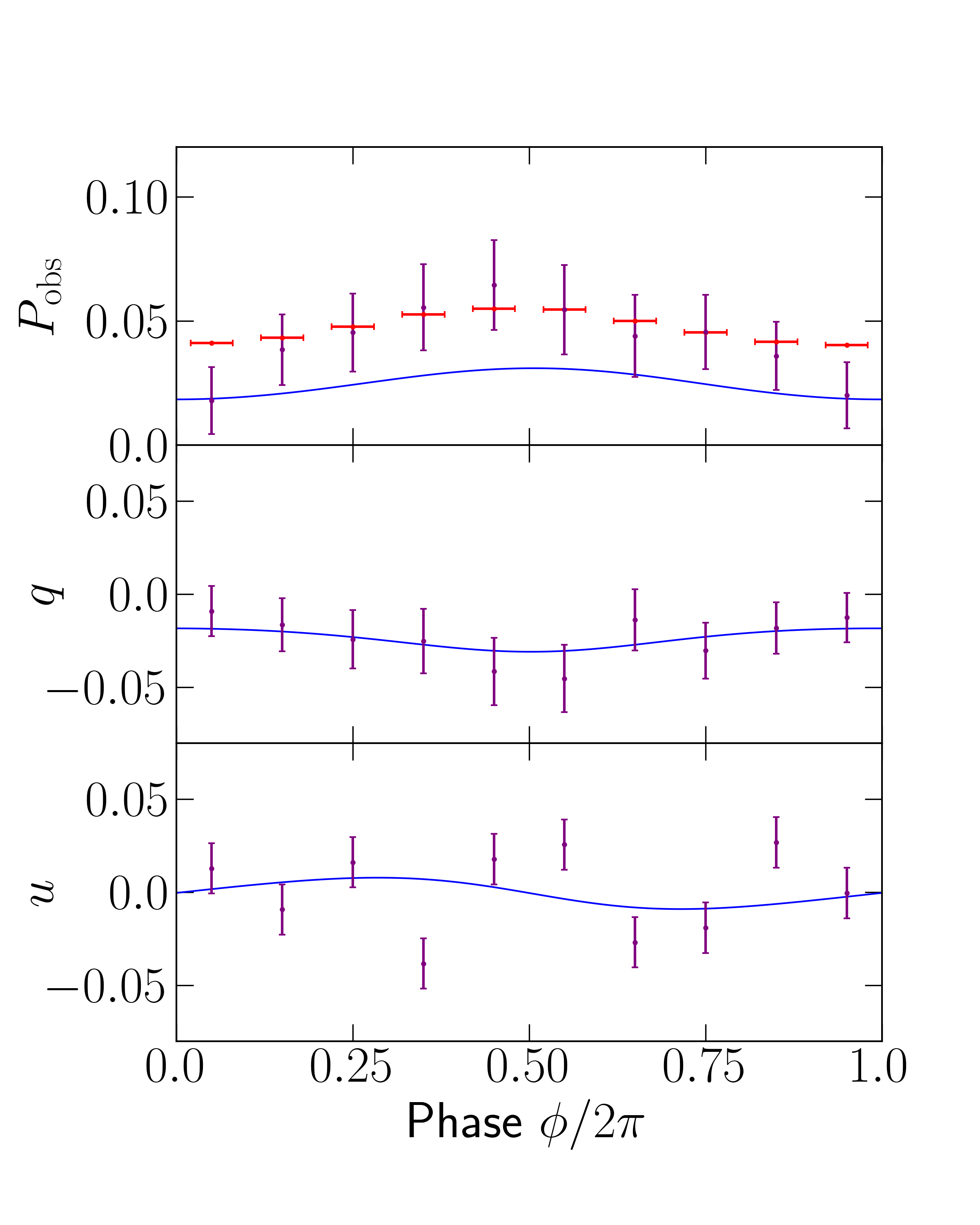}    
\includegraphics[
width=0.49\textwidth]
{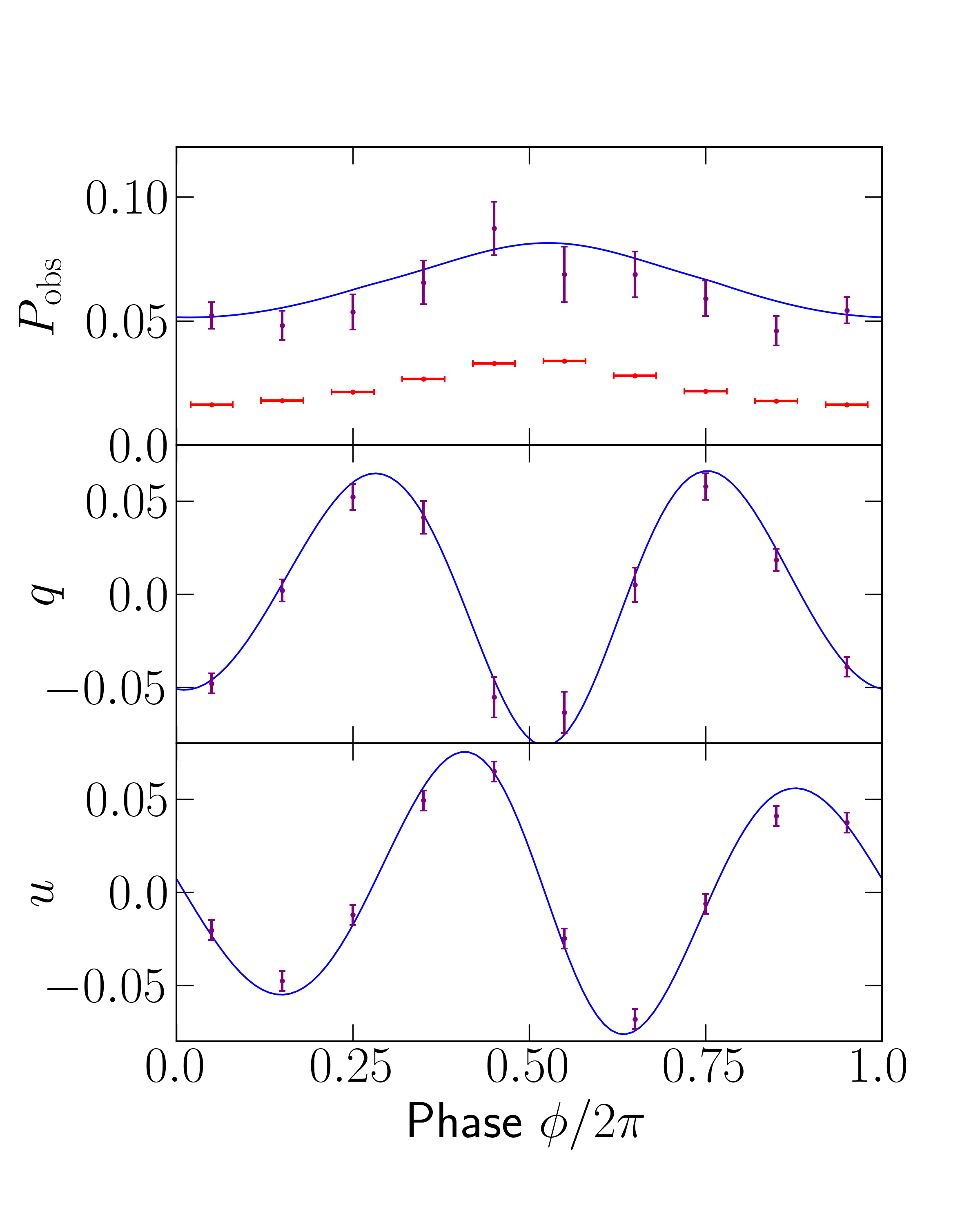}   
\includegraphics[
width=0.49\textwidth]
{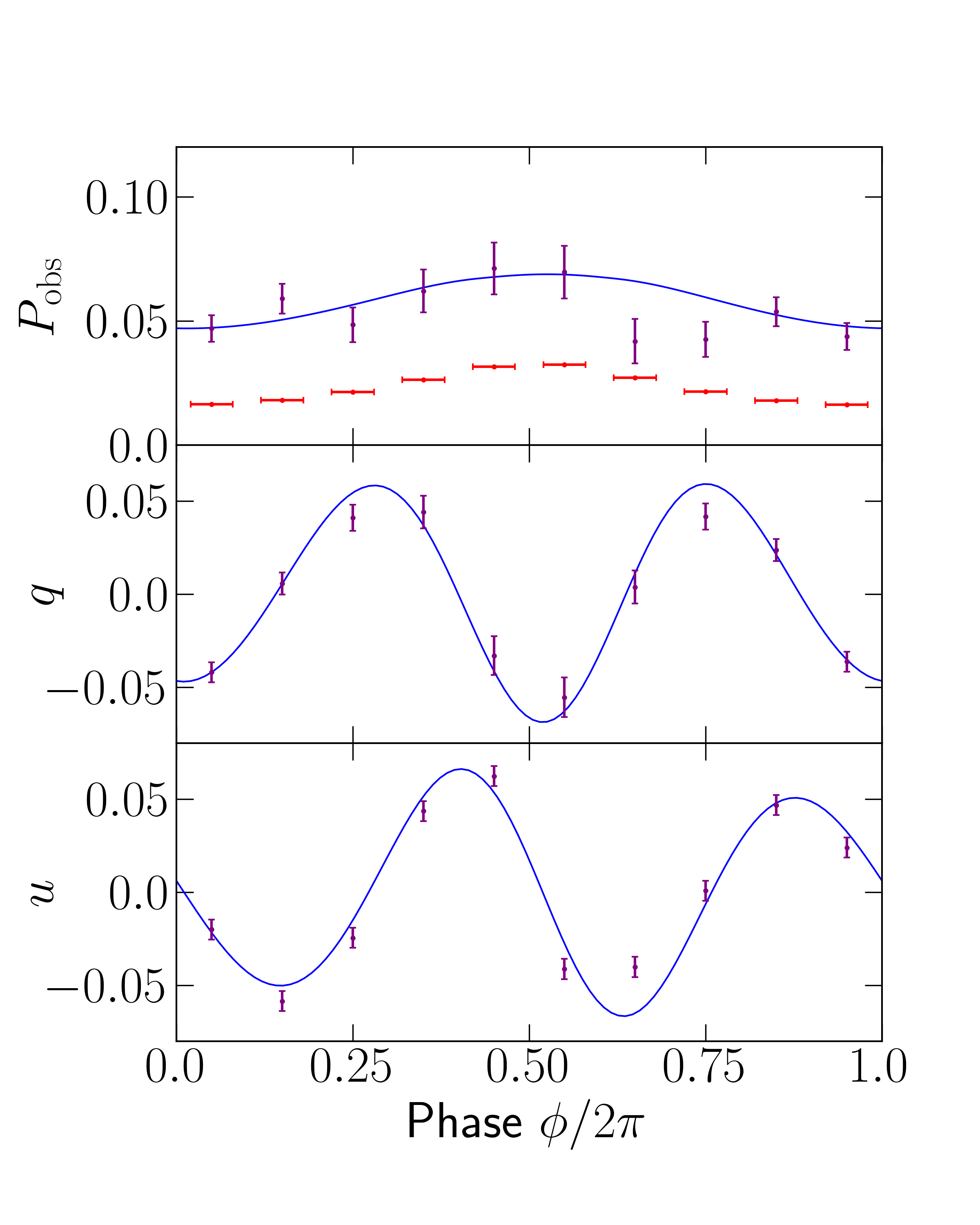}   
\caption{\small{
Synthetic PD ($P_{\mathrm{obs}}$) and normalized Stokes $q$, $u$ data for scenarios A (top left), B (top right), C (bottom left), and D (bottom right). See Section~\ref{sec:input_models} for the input model explanation. The blue curves show the model curve using the injected parameters, the purple dots and error bars show the synthesized observed data, and the red bars show the MDP values (see Section \ref{sec:simulated_data} for a definition) for the corresponding phase points.
}}
\label{fig:synt_data}
\end{figure*}
}

{
\begin{figure*}
\centering
\includegraphics[
width=0.49\textwidth]
{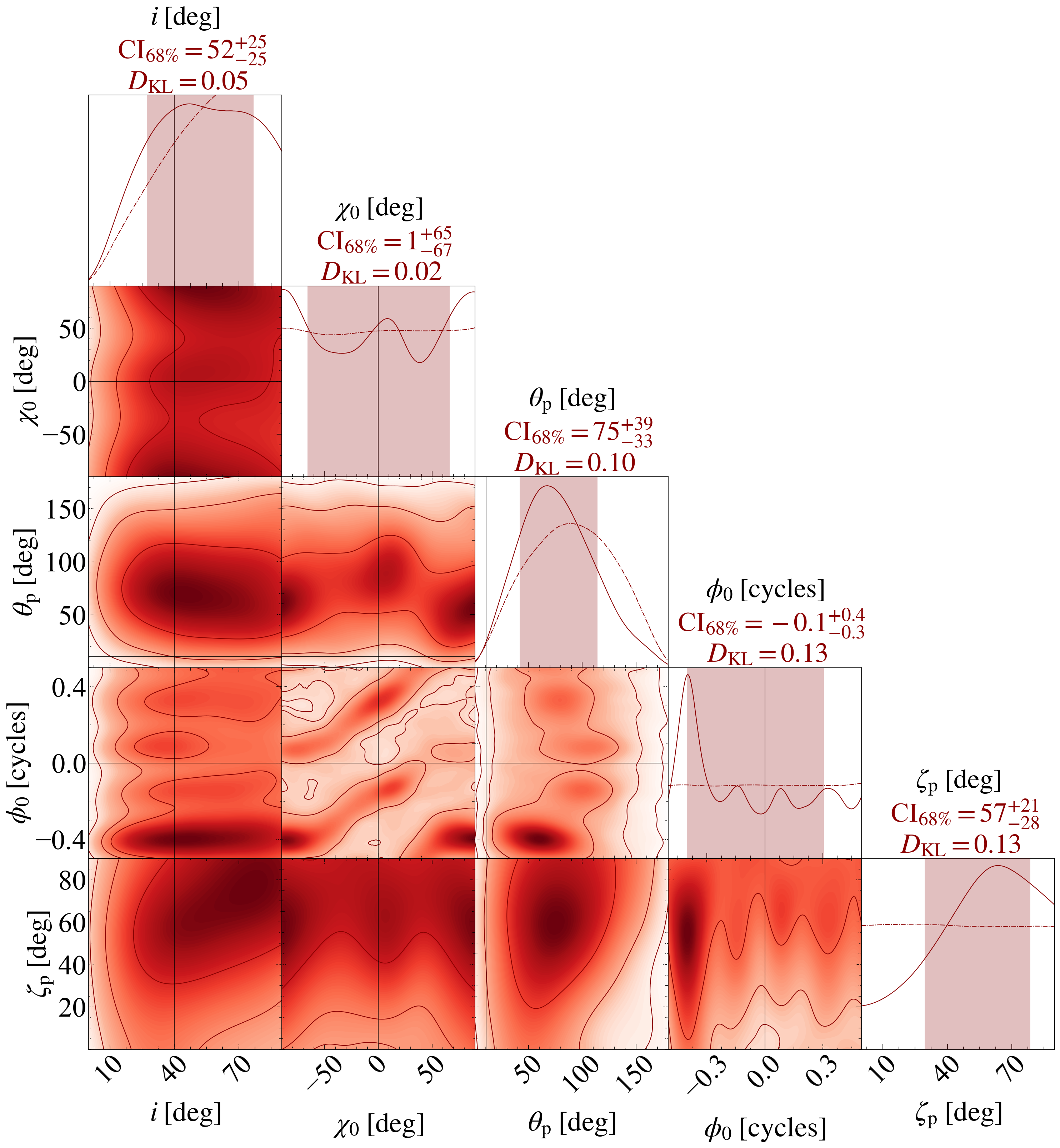}
\includegraphics[
width=0.49\textwidth]
{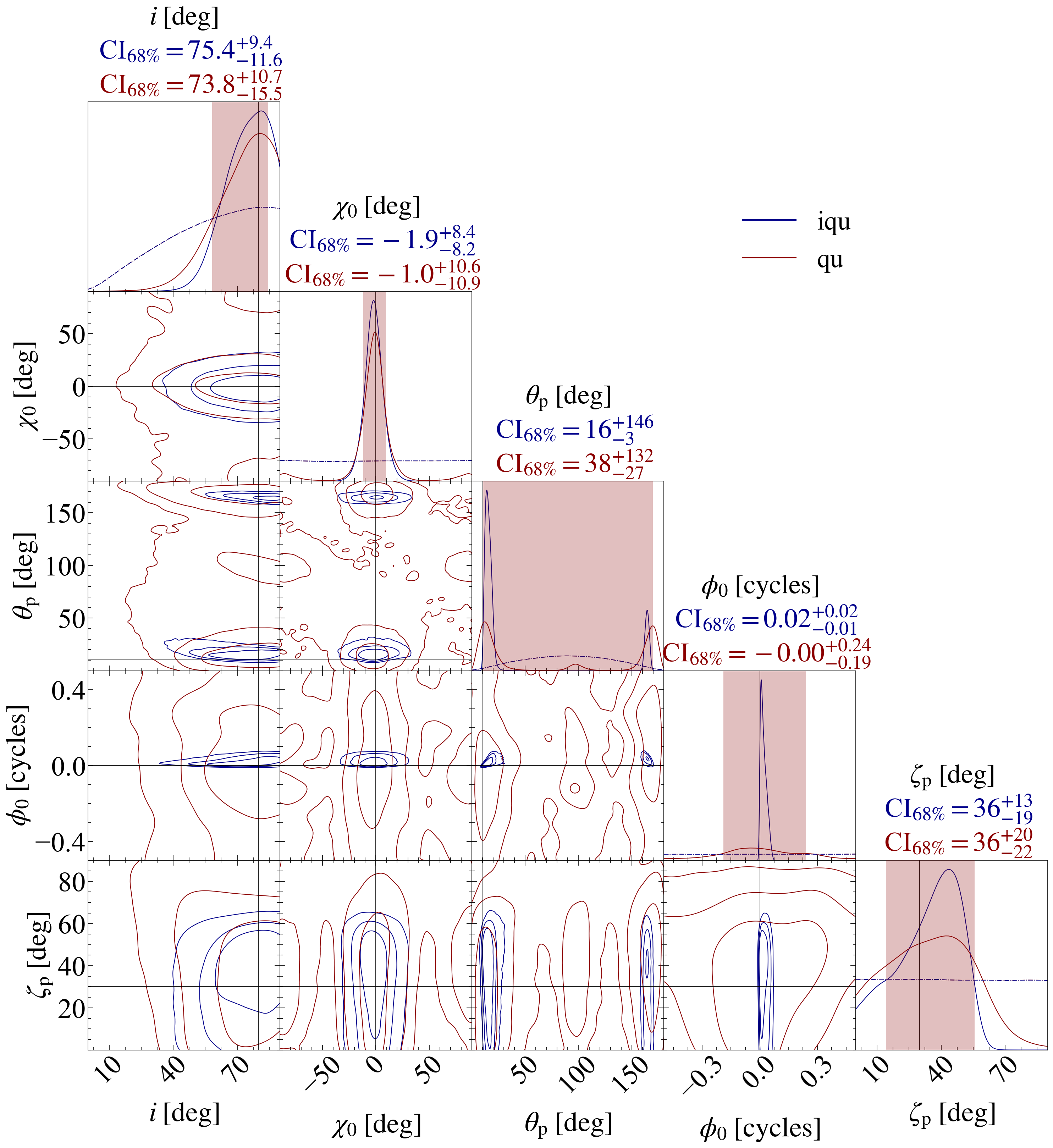}    
\includegraphics[
width=0.49\textwidth]
{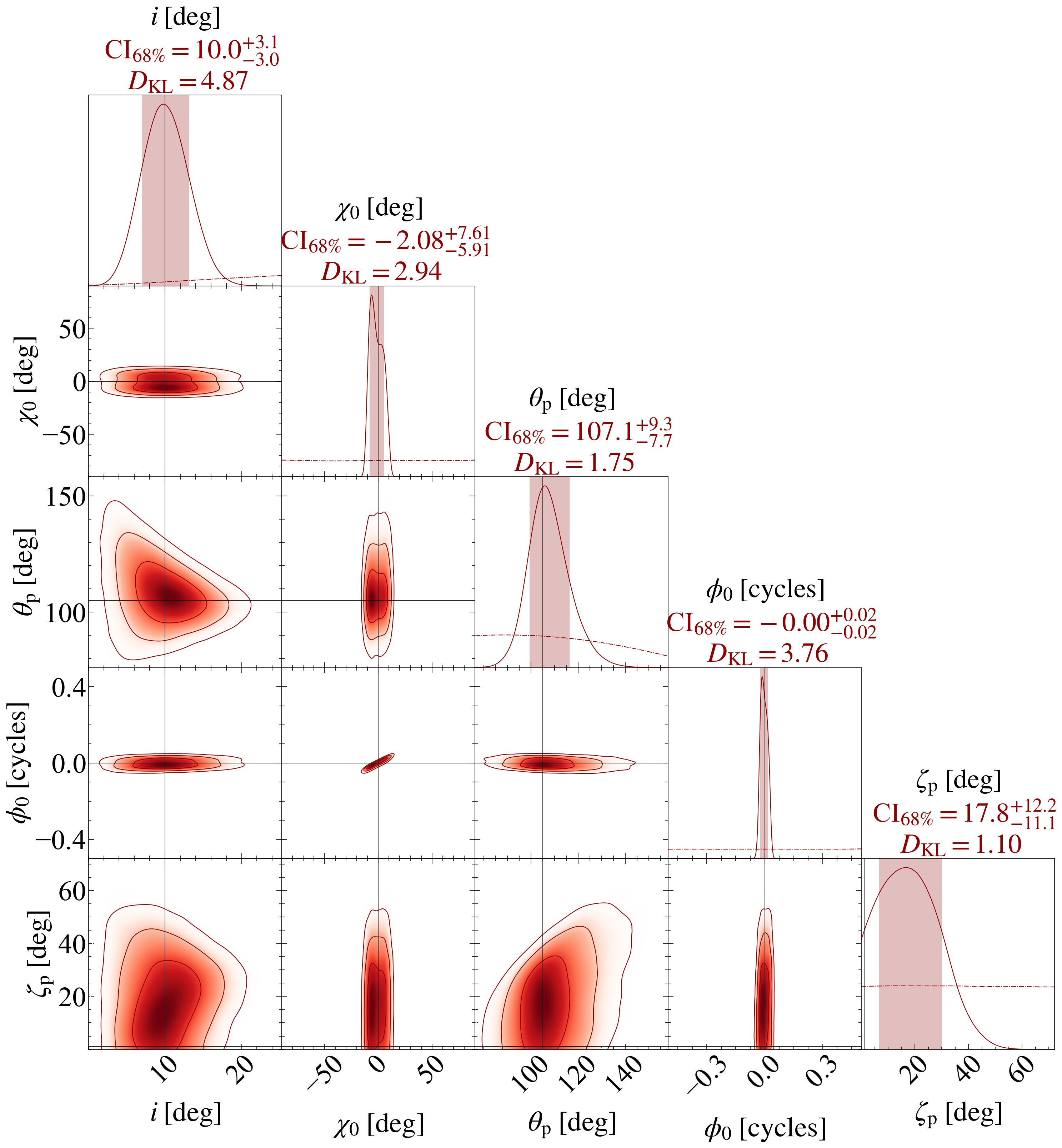}   
\includegraphics[
width=0.49\textwidth]
{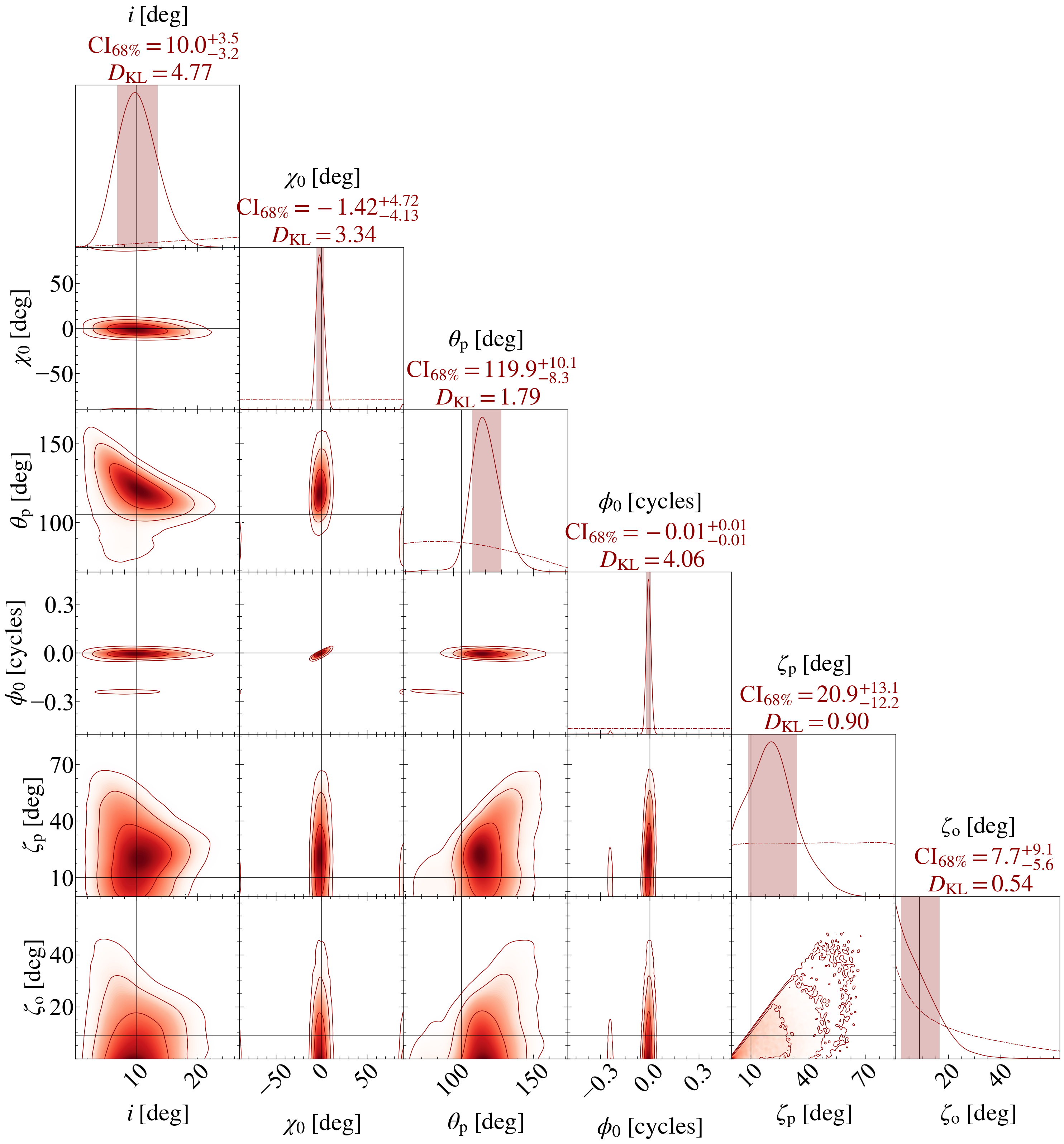}   
\caption{\small{
Posterior distributions of the geometry parameters for scenarios A (top left), B (top right), C (bottom left), and D (bottom right). See Section \ref{sec:input_models} and Table \ref{table:input_params} for the input model definitions.
For Scenario B, results with Stokes $i_{\mathrm{N}}$ included in the analysis are shown (iqu) in addition to those without it (qu).
The dash--dotted curves represent the prior distributions and the thin solid lines show the true values.
The 1D shaded intervals contain $68.3$ per cent of the posterior mass, and the 2D contours contain $68.3$, $95.4$, and $99.7$ per cent of the posterior mass.
KL-divergence values ($D_{\mathrm{KL}}$), describing the prior-to-posterior information gain, are also reported for each parameter (measured in bits as explained in A.2.4. in \citealt{RWB_nicer19}), except in Scenario B where the $D_{\mathrm{KL}}$ values for ($i$, $\chi_{0}$, $\theta_{\mathrm{p}}$, $\phi_{0}$, $\zeta_{\mathrm{p}}$) are ($0.69$, $2.31$, $4.69$, $4.07$, $0.70$) in case iqu and ($0.43$, $1.66$, $2.77$, $0.26$, $0.27$) in case qu.
}}
\label{fig:posteriors_geom}
\end{figure*}

\subsection{Simulated data}\label{sec:simulated_data}

We start by presenting the simulated data and some of its properties. 
All the simulated data sets are visualized in Fig.~\ref{fig:synt_data}.
We see that the expected PD and the measurement uncertainties vary significantly between the different scenarios.
Scenario A leads to practically zero PD, because the polarization is
averaged out across the large hot spot surface.
The data effectively inform only an upper limit for the PD and for the variation in Stokes $q$ and $u$ parameters.
With a smaller hot spot in Scenario B, the PD is notably larger, but still mostly below the minimal detectable polarization (MDP) at 99 per cent significance (see \citealt{WEO10} and \citealt{KCB15} for details about MDP).
However, some information from polarization may still be extracted as we directly fit $q$ and $u$ rather than PD when aiming to constrain the geometry of the source.

Notably tighter measurements of the polarization are obtained in Scenarios C and D. 
In both cases PD exceeds 5 per cent at all phases, and enough photons are observed to constrain the PD, $q$, and $u$ values at each bin with roughly 20 per cent accuracy.
The higher PD is a result of viewing the rather small hot spot always from a direction that requires high emission angles, in which the photons are more polarized.
We see no big differences between the two scenarios (the PD is only slightly smaller for the latter), implying that having circular (Scenario C) versus ring-like (Scenario D) hot spot shape has only a very small influence on the data.

\subsection{Parameter constraints}\label{sec:parameter_constraints}

Next, we present the posterior distributions for the geometry parameters in Fig.~\ref{fig:posteriors_geom} for the same four scenarios introduced in the previous sections.
We see, as expected, that very little or no information about the model parameters can be obtained in Scenario A with a very large hot spot. 
The posterior distributions follow mostly the prior distributions, and the Kullback–Leibler (KL) divergence $D_{\mathrm{KL}}$ (representing prior-to-posterior information gain) is always close to zero.
Only some of the smallest spot angular radii seem slightly disfavoured as those would often produce a higher PD than the observed upper limit.

Notably better parameter constraints are obtained in Scenario B, where we also checked the effect of including Stokes $i_{\mathrm{N}}$.
The inferred inclination angle is $i = (74_{-16}^{+11})\degr$ without $i_{\mathrm{N}}$ and $i = (75_{-12}^{+9})\degr$ with $i_{\mathrm{N}}$, both reported as the 68\% credible intervals around the median.
The corresponding spin axis position angle is $\chi_{0} = -1\degr\pm11\degr$ without $i_{\mathrm{N}}$ and $\chi_{0} = -2\degr\pm8\degr$ with $i_{\mathrm{N}}$.
The injected values ($80 \degr$ and $0 \degr$) are within these intervals.
Significant information gain over the prior is obtained also for the spot colatitude $\theta_{\mathrm{p}}$, although the posterior distribution is strongly bimodal as the signal is very similar regardless of whether the spot is close to the northern or southern rotational pole. 
Thus, the 1D credible interval around the median extends almost over the full prior range.
However, we checked with additional runs that if the $\theta_{\mathrm{p}}$ upper limit had been set to $90\degr$ (e.g. as might be expected if the view of the  opposite hemisphere were blocked by the accretion disc), the credible interval would have been $\theta_{\mathrm{p}} = (14.2_{-6.6}^{+9.5})\degr$ without $i_{\mathrm{N}}$ and $\theta_{\mathrm{p}} = (13.5_{-2.6}^{+3.8})\degr$ with $i_{\mathrm{N}}$ (the true value being $10\fdg3$).
Including $i_{\mathrm{N}}$ provided thus a notable further constraint on the spot co-latitude.
It also significantly helped to constrain the phase zero $\phi_{0}$ as seen in Fig.~\ref{fig:posteriors_geom}.

A significant improvement in all the geometry parameter constraints can be seen for Scenarios C and D.
In both cases the observer inclination is constrained to be around the true value with $i = 10\degr\pm3\degr$.
The hot spot colatitude is also quite tightly inferred, $\theta_{\mathrm{p}}=(107_{-8}^{+9})\degr$ in Scenario C and  $\theta_{\mathrm{p}}=(120_{-8}^{+10})\degr$ in Scenario D.
The true value ($105 \degr$) is outside of the 68 per cent interval in the latter case, but is still within the 95 per cent interval (not shown in the figure).
The exact shape and the size of the hot region is not particularly well constrained in either case.
The primary angular radius is somewhat overestimated in Scenario C ($\zeta_{\mathrm{p}} = (18_{-11}^{+12})\degr$, while the true value is $\zeta_{\mathrm{p}} = 1 \degr$) but falls within the 68 per cent interval in Scenario D ($\zeta_{\mathrm{p}} = (21_{-12}^{+13})\degr$, while the true value is $\zeta_{\mathrm{p}} = 10 \degr$).
In the latter case, the angular radius of the masking circle (i.e. the inner radius of the emitting ring) is only slightly more constrained than its prior distribution with $D_{\mathrm{KL}} = 0.54$, $\zeta_{\mathrm{o}} = (7.7_{-5.6}^{+9.1})\degr$, and the true value $\zeta_{\mathrm{o}} = 9\degr$.
We also checked that analysing the Scenario D data (an emitting ring) using the Scenario C model (an emitting circle) yields an $\sim 5\degr$ larger spot radius, but otherwise very similar parameter constraints, and an insignificant difference in the Bayesian evidence (the ring model being better only by $0.04$ in ln-units).
This demonstrates that these polarization data are not very sensitive to the exact NS surface pattern.

Unlike what has usually been done in previous studies, we kept many of the model parameters other than the geometry parameters free when fitting the simulated data (the NS mass, radius, and spectral parameters).
However, even for the most optimistic Scenarios C and D, we only obtained weak constraints on some of these additional parameters.
The tightest, but still weak, constraints were obtained in Scenario B when fitting also Stokes $i_{\mathrm{N}}$ (see Fig.~\ref{fig:posterior_other_params}).
This is not surprising since the polarimetric data depend only very weakly on these parameters.
Nevertheless, we checked in Scenario B that fixing $T_{\mathrm{seed}}$, $T_{\mathrm{e}}$, $\tau$ and $N_{\mathrm{H}}$ does not substantially affect the computation time or the posteriors of the other parameters.

{
\begin{figure}
\centering
\includegraphics[
width=0.49\textwidth]
{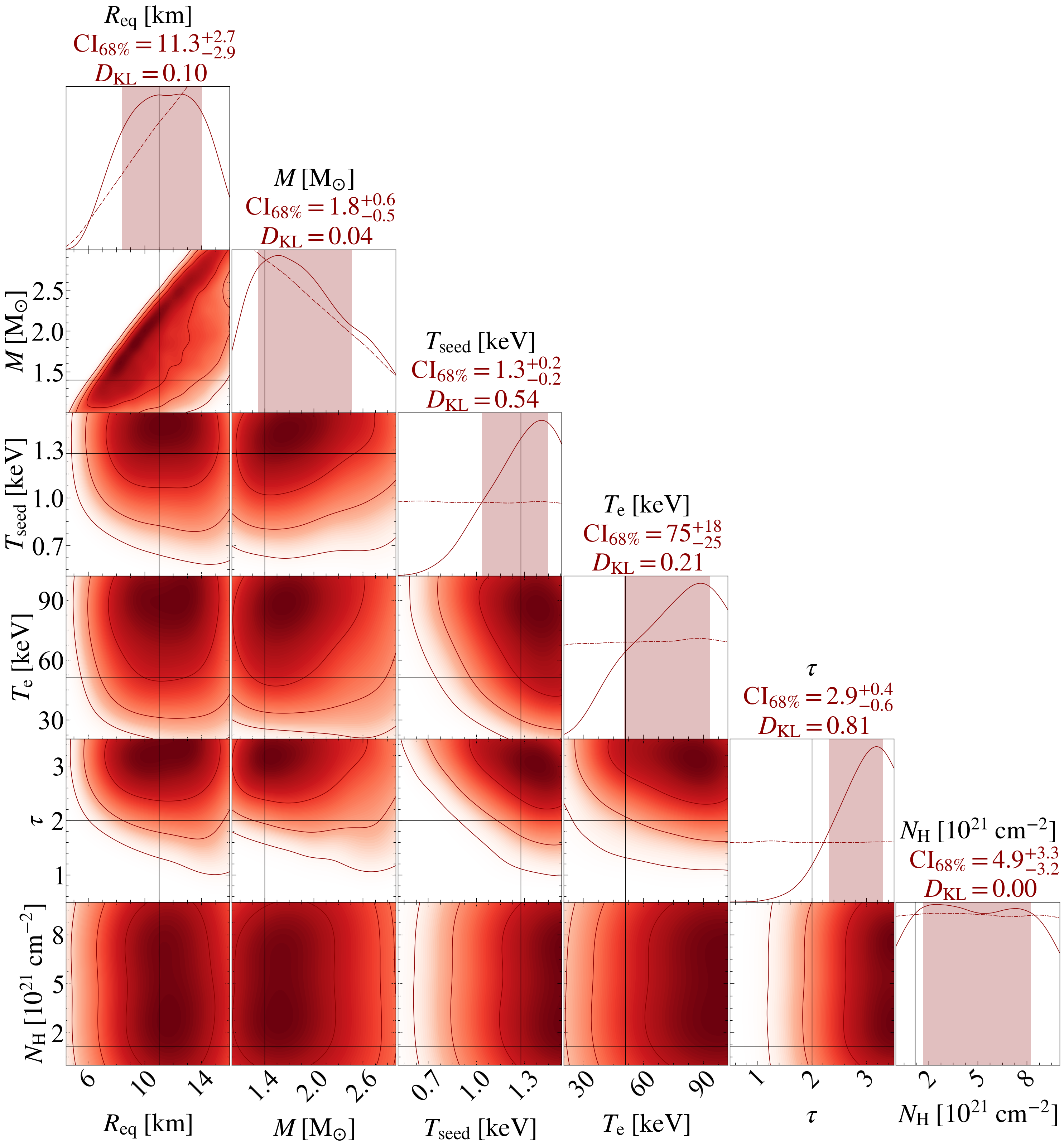}
\caption{\small{
Posterior distributions of the remaining free parameters for Scenario  B (in case iqu). 
See the caption of Fig.~\ref{fig:posteriors_geom} for details of the figure elements.
}}
\label{fig:posterior_other_params}
\end{figure}
}

\section{Discussion}\label{sec:discussion}

Our study shows that the achievable AMP parameter constraints from X-ray polarization depend significantly on the properties of the observed NS, as seen also in \citet{SLK21} and \citet{Bobrikova2023}.
In addition, we found that the NS mass and radius cannot be constrained in any of our scenarios if using only simulated Stokes $q$ and $u$ data.
This is not surprising because the polarization profiles are only weakly dependent on these parameters through the relativistic effects on PA \citep{poutanen20,LSNP20}.
Including the bolometric Stokes $i_{\mathrm{N}}$ (observed by IXPE) in the modelling did not significantly improve the mass--radius constraints in our test case.
However, more accurate results are expected when combining the polarimetry information from IXPE with modelling of energy-resolved pulses observed by X-ray instruments that are focused on collecting high number of photons without polarization information.
This is because the polarimetric information helps to break the degeneracy between the mass and radius and the NS geometry \citep{VP04}.

The current observations of SRGA J144459.2$-$604207 \citep{Ng2024,Molkov2024} have already proved that IXPE is capable of measuring polarization from AMPs  \citep{papitto25}.
The data of this source most resemble our Scenario B and were used to infer (with a simplified model and $\chi^{2}$ fitting) an inclination of $(74.1_{-6.3}^{+5.8}) \degr$, and two small hot spots located at around $(11.8_{-3.5}^{+2.5})\degr$ and $(172.6_{-1.0}^{+2.0})\degr$ colatitudes. 
The method presented here can be used to analyse these data more thoroughly and applied straightforwardly for a joint pulse profile analysis between different instruments (Dorsman et al. in preparation).

Magnetohydrodynamic simulation of accreting stars have shown that hot spot shapes are likely non-circular \citep{Romanova2012,Das2024}.
The shape of the spot is expected to be ring-like for small spot colatitudes but more crescent-like or an elongated band for higher colatitudes \citep{Das2024}.
Therefore, we studied also the effect of different surface patterns on the polarization signal.
However, instead of a ring close to the pole, we considered a ring close to the equator (to maximize PD and to have still pulsations).
Nevertheless, we found that the simulated polarization data were not able to confidently distinguish a ring from a circle. 
But we note that the spot shape could be more important in some other configurations, for example, with larger or differently shaped spots, or when not looking at the spot always from the edge (although this would mean a lower PD) or with data from future instruments.

There are also a few other caveats related to the results presented here.
First, we only consider one hot spot, which is justified if the accretion disc blocks the view of a secondary hot spot \citep{IP09}.
The effects of including two hot spots were tested by \citet{SLK21} who found that the location of both of them could be constrained using polarimetry.
Secondly, the model used in this paper can be developed further by replacing the isothermal slab approximation with a model where dynamics of the accreting gas is computed self-consistently with the electron temperature.
Finally, the effects on polarization due to the scattering in the accretion column could also be relevant, but accounting for that would need a more comprehensive radiative transfer model \citep[see][for the effects on non-polarized pulses]{Ahlberg2024,Das2024}.
However, the IXPE data, especially in Scenarios like A and B, are not expected to be very sensitive to these effects.
We note that slightly more information of the model parameters could be obtained by modelling the non-normalized Stokes $F_{Q}$ and $F_{U}$ parameters with an energy-resolved forward-folding approach (with an increase in the computational expense).
This is also possible with the \XPSI pipeline developed here, but applying this approach is left for future work.

\section{Conclusions}\label{sec:conclusions}

We simulated X-ray polarization data and estimated NS parameters constraints for AMPs with different hot spot and data quality scenarios.
We used similar techniques as in \citet{Bobrikova2023}, but now implemented in \XPSI, allowing a straightforward exploration of more complex hot spot shapes and joint analyses with pulse profile data from different instruments like NICER and \textit{XMM-Newton}. 
We found that the observer viewing angle and the hot spot location on the surface can be well constrained in some scenarios, but not if the emitting region is very large and the PD very small.
Our current simulations were not able to robustly distinguish between two different hot spot shapes (a circle versus ring).
However, the already existing AMP X-ray polarization observations by IXPE \citep{papitto25} demonstrate that NS geometry, and thus also mass and radius, will likely be further constrained with the help of X-ray polarimetry.
Detailed studies of these data are in progress.

\section*{Acknowledgements}
TS, BD, and ALW acknowledge support from European Research Council (ERC) Consolidator Grant (CoG) No.~865768 AEONS (PI: Watts).
ADM is supported by the Italian Space Agency (Agenzia Spaziale Italiana, ASI) through contract ASI-INAF-2022-19-HH.0.
JP was supported by the Ministry of Science and Higher Education grant 075-15-2024-647.
AP acknowledges support supported by INAF 2022 Research Grant FANS (PI: Papitto), Italian Ministry of University and Research PRIN 2020 Grant 2020BRP57Z, (PI: Astone) and Fondazione Cariplo/CDP, grant no. 2023-2560 (PI: Papitto).
The use of the national computer facilities in this research was subsidized by NWO Domain Science.
This work used the Dutch national e-infrastructure with the support of the SURF Cooperative using grant no. EINF-8005.  

\section*{Data availability}
A basic reproduction package including complete information for each run, simulated data, posterior sample files, and the analysis scripts is found in the Zenodo repository at doi:\href{https:/doi.org/10.5281/zenodo.14130094}{10.5281/zenodo.14130094}.

\bibliographystyle{mnras}
\bibliography{allbib}

\label{lastpage}

\end{document}